\documentclass[11pt]{article}  
\usepackage[utf8]{inputenc}  
\usepackage[T1]{fontenc}  
\usepackage{textcomp}  
\usepackage{lmodern} 
\usepackage{amsmath, amssymb}  
\usepackage{graphicx}
\usepackage[a4paper, margin=1in]{geometry} 
\usepackage{setspace} 
\usepackage{titling} 
\usepackage{hyperref}

\usepackage{comment}

\usepackage[scaled]{helvet}

\usepackage{tikz}

\usepackage{pdflscape}  
\usepackage{booktabs}  
\usepackage{threeparttable}

\usepackage{graphicx}

\usepackage{array}

\usepackage{xcolor}        
\usepackage{hyperref}      
\usepackage{graphicx}      
\usepackage{caption}       
\usepackage{booktabs}      

\hypersetup{
    colorlinks=true,       
    linkcolor=blue,        
    citecolor=blue,        
    filecolor=blue,        
    urlcolor=blue          
}

\usepackage[authoryear]{natbib}
\usepackage{bibentry} 

\begin{document}


\begin{titlepage}
    \begin{center}
        \vspace*{1in}

        {\LARGE \textbf{Using Covid-19 Response Policy to Estimate Open Water and Swim Drafting Effects in Triathlon}}\\[1.5cm]

        \Large{Felix Reichel}\\[10cm]

        \large{Corresponding Author:}\\
        \large{Felix Reichel}\\
        \large{eMail: \url{felix.reichel@jku.at}}\\[1.5cm]


        \large{Word Count: $\approx$ 2830 words} \\[0.5cm]



        \vfill
    \end{center}
\end{titlepage}

\begin{flushbottom}
    \noindent \footnotesize EU-GDPR compliant anonymized data sponsoring by Triathlon Statistics Austria and Mag. Thomas Doblhammer is gratefully acknowledged.
\end{flushbottom}


\onehalfspacing

\newpage



\newpage
\section*{Abstract}

\subsection*{Objectives}
This study investigates the causal effects of open-water swim drafting by leveraging a natural experiment induced by staggered race starts during the COVID-19 pandemic. Furthermore, the study aims to quantify the benefits of swim drafting in real-world triathlon race settings and estimate the effect of different drafting positions within swim groups.

\subsection*{Design}
A quasi-experimental design is employed, exploiting exogenous variation in swim drafting opportunities caused by COVID-19 restrictions. The study uses panel data from triathlon races before, during, and after the pandemic.

\subsection*{Methods}
Using agglomerative hierarchical clustering, likely swim group formations are computed from swim-out segment times. The impact of drafting positions on swim performance is estimated using a two-way fixed-effects regression model, controlling for athlete- and event-specific factors. Robustness checks in the form of alternative regression model specifications and samples are used to partially address some potential endogeneity concerns, including reverse causality and omitted variable bias.

\subsection*{Results}
Empirical findings reveal that drafting benefits were statistically insignificant in 2020 due to reduced drafting opportunities but re-emerged post-pandemic at slightly lower levels than pre-pandemic periods. The results indicate that drafting only becomes beneficial from the third trailing position onward, with earlier positions primarily serving to minimize fatigue, as shown by an observed inverse decay in drafting benefits (higher estimated penalties in lower-order drafting positions/stronger higher-order drafting effects).

\subsection*{Conclusions}
This study provides the first large-scale causal estimate of open-water swim drafting effects in real triathlon race conditions. The findings contribute to the understanding of race strategy optimization (metabolic costs), offering new insights for endurance athletes, race organizers and regulation authorities. Additionally, the results indicate the importance of strategic positioning within swim groups to maximize performance benefits while mitigating fatigue effects.

\newpage

\section*{Introduction}

Aerodynamics plays a significant role in everyday life, whether you own a car, book a flight, or race your neighbor in a cycling competition. In cycling, rotational overtaking maneuvers and shared drafting advantages have long been a common practice, both in racing and training. Similarly, in triathlon, the role of drafting is well recognized, leading to the existence of two distinct race formats:

\begin{enumerate}
    \item \textbf{Elite short- and middle-distance races}, where wind-drafting is explicitly allowed, transforming these events into highly strategic competitions.
    \item \textbf{Long-distance and amateur triathlon races}, where wind-drafting is strictly prohibited, and violations result in time penalties.
\end{enumerate}

In these latter formats, the only real opportunity for drafting occurs in the \textbf{swim segment}, where swimmers can strategically position themselves to reduce drag. Swim drag reduction through drafting has been extensively studied in medical science, soft matter physics and sports science in \citet{Bolon2020, Chatard1990, ChatardWilson2003, Ohmichi1983, Millet1999, Delextrat2003, Olbrecht2011, Toussaint1989}. However, \textbf{no large-scale causal estimates of its benefits in real race settings have been published}.

The \textbf{COVID-19 pandemic} introduced precautionary measures in Austria which significantly altered the swim segment dynamics. According to \citet{Triathlon2020} Athletes often had to start with individually staggered swim entries, and overtaking at buoy turns was sometimes somewhat restricted. This created a \textbf{natural experiment} where COVID-19 functioned as an \textbf{exogenous short-term shock (1–3 years)} to open-water swim drafting, providing an opportunity to analyze its effects in a controlled manner.

In this paper, I employ \textbf{agglomerative hierarchical clustering} of \textbf{swim-out times} to infer likely \textbf{swim group formations} and estimate the optimal drafting positions within swim-groups in race settings. This is the first study to provide \textbf{large-scale causal estimates for open-water  drafting effects using panel data}.

Water drafting has long been a \textbf{hot topic} in both sports science and professional triathlon. With internet sources claiming that an athlete can gain up to between $\approx$ 7\%-21\% on overall drag reduction, which results in a reduction of 2 minutes over a half-distance Ironman swim of 1,900m.\footnote{Drafting in Triathlon: \url{https://support.myprocoach.net/hc/en-us/articles/360022528551-Drafting-in-Triathlon\#:~\%3Atext=Swim\%20Drafting,you\%20up\%20to\%20two\%20minutes.}
}

Beyond race performance, other strategic considerations—such as \textbf{energy conservation} and positioning to \textbf{catch the first cycling group} in elite competitions where wind drafting is permitted—play a critical role.


\newpage

\section*{Related Literature}

Empirical research on open-water swim drafting has consistently demonstrated its benefits in terms of reduced drag, energy conservation (metabolic costs), and improved race performance. \\

\noindent \textbf{Hydrodynamic and Physiological Benefits of Drafting.}
Studies have shown that drafting significantly reduces drag and physiological strain. \citet{Bolon2020} determined that optimal drafting positions—directly behind a lead swimmer or at the hip—result in drag reductions of 40\% and 30\%, respectively. \citet{ChatardWilson2003} found that metabolic costs, including oxygen uptake, heart rate, and blood lactate levels, were significantly lower when swimmers drafted at distances of 0 to 50 cm behind another swimmer's toes. The most advantageous drafting distances were 0 and 50 cm, reducing passive drag by 21\% and 20\%, respectively. Additionally, \citet{Ohmichi1983} found that swimmers drafting directly behind another competitor completed 400-m races significantly faster than those who swam at the side or hip.\\

\noindent \textbf{Drafting and Performance Prediction.}
The role of passive drag (Dp) in swim performance prediction was examined by \citet{Chatard1990}, who found a strong correlation between Dp and VO$_2$ max. Performance times were primarily influenced by VO$_2$ max ($r=0.70$ for males, $r=0.72$ for females), but including Dp significantly improved the predictive accuracy. The strong correlation hints at the possibility that more able athletes might be more sensitive to properly utilizing passive drag. Other anthropometric factors, including height, weight, and body surface area, also contributed to drag variability. This athlete heterogeneity underscores the need to include athlete-specific intercepts or characteristics in predictive models. \citet{Millet1999} further emphasized the cumulative impact of drafting, showing that triathletes employ energy-conserving strategies in swimming to enhance their subsequent cycling and running performances.\\

\noindent \textbf{Tactical Implications for Triathletes.}
Drafting has been shown to enhance subsequent cycling performance. \citet{Delextrat2003} reported a 4.8\% improvement in cycling efficiency when a swim was performed in a drafting position compared to an isolated swim. \citet{Millet1999} also stated the cumulative effects of drafting in triathlon, where athletes strategically conserve energy in the swim segment to optimize performance in the cycle and run segment. \citet{Olbrecht2011} thus underscores the importance of balancing conditioning and technique, arguing that poor swimming technique results in excessive energy expenditure, ultimately affecting overall race outcomes.\\

\noindent \textbf{Impact of Equipment and External Conditions.}
The effect of wetsuits on swim drafting was investigated by \citet{Toussaint1989}, who observed a 14\% reduction in drag at typical triathlon swim speeds. This reduction, largely attributed to increased buoyancy and decreased frontal resistance, resulted in higher swim velocities. Moreover, \citet{Leitner2021} conducted an athlete survey revealing preferences regarding mass starts versus staggered starts, highlighting the impact of the race format on drafting opportunities. An alternative survey on wetsuit use could further reveal preferences for drafting but is likely highly confounded by the swim-to-bike transitional race segment. The COVID-19 pandemic introduced natural experiments with staggered starts, as documented by \citet{Triathlon2020}, which temporarily altered swim drafting dynamics. \\ \\

\noindent \textbf{Summary of Existing Literature.}
The body of research highlights drafting as a critical factor in competitive swimming and triathlon. Hydrodynamic studies focus on reductions in drag, while physiological research conclude that drafting lowers metabolic costs. Performance predictions could therefore benefit from incorporating passive drag measures. External conditions, such as wetsuit use (unobservable, if allowed in an event and race format subject to race organizers and race regulation authorities), technological wetsuit evolution and innovative race formats (such as the 'Super Sprint' e.g.) or short-term race format shocks (such as the Covid-19 pandemic), further influence drafting effectiveness.

\newpage

\section*{Data}

The dataset originates from three relational database tables  provided by the Triathlon Statistics Austria database, covering most official triathlon races that took place in Austria from 2010 to 2024. 

Initially, the dataset contained 175,740 recorded observations. After removing \textbf{DNF} (Did Not Finish) and \textbf{DNS} (Did Not Start) outcomes, as well as missing data, I obtained a final dataset of 168,391 observations.

\subsection*{Dataset Structure}

The dataset is structured into three relational tables:

\begin{itemize}
    \item $\textbf{athlete}(a)$: Contains athlete-specific a details.
    \item $\textbf{events}(e, t)$: Represents official triathlon events e over time t.
    \item $\textbf{result}(r, a, e, t)$: Stores results r of each athlete a in specific events e captured at time t.
\end{itemize}

%
%

\subsection*{Preprocessing and Period Definitions}

The dataset was processed by first merging the result table with the events and athlete table using foreign keys. Events were then assigned to one of three periods based on their date, resulting in three binary dummy variables. Additionally, event-time-variant covariates such as age or squared age were computed in the final dataset as \( \text{age}_{a,t} := \text{year}_e(t) - \text{year}_a \), which were then additionally used as a distributional covariate balance check across periods.

\subsection*{Construction of Swim Groups}

To analyze drafting effects, swimmers were clustered based on their swim-out time within an event. A hierarchical clustering algorithm was applied with a 5-second threshold for forming swim groups.

\textbf{It is noted that in the initial model trials, clustering was also performed using the complete linkage method with an increased cutoff threshold of 15 seconds before finalizing the single linkage approach.}

The clustering procedure involved calculating the time difference between athletes in an event and grouping them based on proximity. A single linkage method was used, ensuring that swimmers within 5 seconds of each other were assigned to the same swim group.

Each cluster was assigned a leader, defined as the athlete with the fastest swim-out time. All other athletes in the group were designated as drafters. Within the drafting group, individual positions were encoded, categorizing athletes as first, second, third, fourth, fifth, and/or last drafters based on their relative swim-out times.

\subsection*{Summary statistics}
\begin{table}[h]
    \centering
    \begin{threeparttable}
        \caption{Summary Statistics of numerical variables}
        \label{tab:summary_stats}
        \begin{tabular}{lcccccc}
            \toprule
            Variable & Min & 1st Quartile & Median & Mean & 3rd Quartile & Max \\
            \midrule
            Total & 165 & 4965 & 8232 & 11789 & 16809 & 61282 \\
            Swim-Out Times & 4 & 825 & 1431 & 1625 & 2097 & 29700 \\
            Race Rank & 0 & 28 & 69 & 156.9 & 151 & 2623 \\
            Male & 0.0000 & 1.0000 & 1.0000 & 0.7898 & 1.0000 & 1.0000 \\
            Year & 1922 & 1969 & 1977 & 1977 & 1985 & 2009 \\
            Covid & 0.0000 & 0.0000 & 0.0000 & 0.1097 & 0.0000 & 1.0000 \\
            Post Covid & 0.00000 & 0.00000 & 0.00000 & 0.08563 & 0.00000 & 1.00000 \\
            Week Running & 0.0 & 159.0 & 279.0 & 317.6 & 475.0 & 751.0 \\
            Age & 8.00 & 31.00 & 38.00 & 38.56 & 46.00 & 99.00 \\
            Age Squared & 64 & 961 & 1444 & 1600 & 2116 & 9801 \\
            Event Year & 2010 & 2013 & 2015 & 2016 & 2019 & 2024 \\
            Cluster (Swim Group) & 1.00 & 4.00 & 10.00 & 16.95 & 23.00 & 224.00 \\
            Leader & 0.0000 & 0.0000 & 0.0000 & 0.3134 & 1.0000 & 1.0000 \\
            Drafter & 0.0000 & 0.0000 & 1.0000 & 0.6866 & 1.0000 & 1.0000 \\
            Drafter Position & 1.00 & 1.00 & 3.00 & 17.15 & 11.00 & 937.00 \\
            First Drafter & 0.0000 & 0.0000 & 0.0000 & 0.2961 & 1.0000 & 1.0000 \\
            Second Drafter & 0.0000 & 0.0000 & 0.0000 & 0.1342 & 0.0000 & 1.0000 \\
            Third Drafter & 0.00000 & 0.00000 & 0.00000 & 0.08329 & 0.00000 & 1.00000 \\
            Fourth Drafter & 0.00000 & 0.00000 & 0.00000 & 0.05846 & 0.00000 & 1.00000 \\
            Fifth Drafter & 0.000 & 0.000 & 0.000 & 0.044 & 0.000 & 1.000 \\
            Last Drafter & 0.0000 & 0.0000 & 0.0000 & 0.2961 & 1.0000 & 1.0000 \\
            \bottomrule
        \end{tabular}
        \begin{tablenotes}
            \item \textit{Notes:} This table presents summary statistics for the key variables used in the analysis. The dataset consists of 168,391 observations. Variables include athlete characteristics, race details, and encoded swim group (cluster), drafting, drafting position binary dummy variables. Encoded categorical variables such as Event Category \(\in \{\text{Sprint}, \text{Short}, \text{Middle}, \text{Long}\} \) are not displayed.
        \end{tablenotes}
    \end{threeparttable}
\end{table}

\newpage

\subsection*{Balance Checks}
To ensure comparability (e.g. problems with the panel data such as attrition) across event categories and periods, some descriptive and visual balance checks on key event, result and athlete variables were performed.

\paragraph{Swim-out times across Event Category and Period Balance} 
Swim-out times were analyzed across different event categories (Sprint, Short, Middle, and Long) and periods (Pre-Covid, Covid, and Post-Covid) to assess differences. The results indicate slight increases in mean swim-out times post-Covid for most Sprint and Short events. This trend however was expected, as the movement towards more accurate swim distances and longer swim distances in the Short category over the last decade is well known among Austrian triathlon insiders. The balance check is shown in table \ref{tab:balance_evn}.

\paragraph{Athlete Characteristics Across Periods} 
Athlete-level characteristics, including mean swim-out time, rank, age, and the proportion of male participants, were examined to assess balance across periods. Age distribution trends indicate a slight increase in the average age of athletes since the pre-Covid period starting in 2010, alongside a slight decline in the overall high proportion of male participants. Additionally, the observed decrease in mean rank suggests a shift in event market structure, potentially favoring a higher number of smaller events or fewer mass events. However, this trend could also be entirely driven by an outlier event within the longer period spanning 2010 to 2019.

    \begin{table}[h]
    \centering
    \caption{Results (Swim-Out Times) across different Event Categories and Periods}
            \label{tab:balance_evn}
    \begin{tabular}{llrrrr}
        \toprule
        Event Category & Period & Mean Swim-Out Time & SD Swim-Out  Time & Observations \\
        \midrule
        Short   & Covid       & 1855  & 380  & 4694  \\
        Short   & Post-Covid  & 1899  & 403  & 3726  \\
        Short   & Pre-Covid   & 1696  & 415  & 37722 \\
        \midrule
        Long   & Covid       & 4594  & 804  & 1130  \\
        Long   & Post-Covid  & 4572  & 1035 & 1014  \\
        Long   & Pre-Covid   & 4553  & 773  & 10201 \\
        \midrule
        Middle & Covid       & 2311  & 380  & 4658  \\
        Middle & Post-Covid  & 2315  & 394  & 3527  \\
        Middle & Pre-Covid   & 2216  & 436  & 26711 \\
        \midrule
        Sprint & Covid       & 854   & 295  & 7985  \\
        Sprint & Post-Covid  & 906   & 296  & 6152  \\
        Sprint & Pre-Covid   & 774   & 258  & 60871 \\
        \bottomrule
    \end{tabular}
\end{table}


\newpage



\section*{Identification Strategy}

\subsection*{Event-Induced Altered Drafting Opportunity.} This study leverages the staggered race starts implemented during the COVID-19 pandemic to estimate the magnitude of drafting benefits in open-water swimming. Before 2020, athletes started in groups, allowing drafting effects to influence race performance. However, during the pandemic, individually staggered starts significantly reduced drafting opportunities. This natural experiment offers a unique opportunity to assess drafting’s impact by comparing race outcomes before, during, and after the pandemic.

\subsection*{Confoundedness, Effect Heterogeneity and Higher-Order Drafting Position Effects.} To formally identify the effects of drafting, I estimate the impact of drafting and within-swim-group position on Swim-Out times, using likely constructed swim groups based on clustered Swim-Out times. Furthermore, I employ athlete and event fixed effects to control for unobservable heterogeneity and to partially address some endogeneity concerns such as omitted variable bias and simultaneous reverse causality. 

The positioning within a swim group likely affects the magnitude of drafting benefits, with higher-order positions benefiting more from overall reduced water resistance (increased passive drag), leading to varying performance outcomes. However, reverse causality is possible if swimmers in the last positions within swim groups only manage to catch up towards the end of the swim segment and are, in general, more skilled and/or performing athletes. Additionally, a strong correlation between passive drag and $\dot{V}O_{2\max}$ could also indicate that stronger athletes also experience greater passive drag \citet{Chatard1990}, assuming this result applies also to real-world race settings.

\subsection*{Unobservables and COVID-19 Event Regulations.} Single catching-up scenarios cannot be fully accounted for in the estimated models. The potential bias introduced by such instances is assumed to be uniformly distributed across all race observations. 

Special consideration is given to regulatory constraints, such as the COVID-19 rule prohibiting overtaking in specific course segments. These restrictions further influence race drafting opportunities throughout the swim segment, ultimately affecting overall mean race outcomes for drafters. Additionally, different observable course configurations, such as triangular and circular swim courses, shape drafting benefits and race dynamics. 

Furthermore, Buoy placements create tactical points that influence overtaking and positioning strategies, particularly in multi-lap races where swimmers navigate the same course multiple times before exiting the swim segment.

\newpage
\section*{Estimation Framework}

To estimate the impact of swim drafting on Swim-Out performance, I employ a Two-Way Fixed Effects (TWFE) model, controlling for both athlete- and event-specific heterogeneity. The general estimation equation can be written in the form of:

\notag \nonumber \begin{equation*}
    y_{aert} = \beta_0 + \beta_1 \text{Leader}_{aert} + \sum_{p=1}^{P} \beta_p \text{Drafter}_{aert}^{(p)} + \boldsymbol{\theta}^\top \mathbf{X}_{aert} + \eta_{et} + \eta_{a} + \varepsilon_{aert}
    \label{eq:base_model}
\end{equation*}

where \( y_{aert} \) represents the Swim-Out performance (e.g., swim time) of athlete \( a \) in event \( e \), stored as race result \( r \), at time \( t \). The encoded binary dummy variables \( \text{Leader}_{aert} \) and \( \text{Drafter}_{aert}^{(p)} \) indicate drafting positions, while the vector \( \mathbf{X}_{aert} \) contains relevant controls, including the overall race rank or constructed cluster rank (Swim Group) in the swim segment.

The fixed effects \( \eta_{a} \) account for athlete heterogeneity, while \( \eta_{et} \) captures event-time-specific shocks such as weather conditions, water temperature, or changes in race regulations. (e.g. wet suit allowance)

To explore period heterogeneity in drafting effects, I estimate a second specification with interaction terms:
\nonumber \begin{equation*}
    y_{aert} = \beta_0 + \text{Drafter}_{aert}  \times (\beta_1 \text{PrePeriod}_t + \beta_2 \text{CovidYear}_t) + \boldsymbol{\theta}^\top \mathbf{X}_{aert} + \eta_{et} + \eta_a + \varepsilon_{aert}
    \label{eq:interaction_model}
\end{equation*}

This model allows us to estimate whether the effect of drafting changed over time, particularly before, during, and after COVID-19, when race regulations varied significantly.







\subsection*{Clustered, Robust, and TWFE Standard Errors}

To ensure reliable statistical inference, I employ various robust standard error estimators that account for heteroskedasticity and correlations within athletes across multiple events.

\subsubsection*{Heteroskedasticity-Robust (HC1) Standard Errors}

To address potential heteroskedasticity—where the variability of errors differs across observations—I use a robust standard error estimator. This approach adjusts the standard errors to remain valid even when the underlying variance structure is inconsistent across data points.

\subsubsection*{Clustered Standard Errors at the Athlete Level}

To account for the fact that individual athletes may participate in multiple events, I cluster standard errors at the athlete level. This method ensures that the estimated errors account for within-athlete correlations, leading to more accurate inference when analyzing repeated observations from the same athlete.

\subsubsection*{Clustered Standard Errors at the Athlete and Event Level}

Since dependencies may arise not only within athletes but also within events, I implement a multi-way clustering approach. This accounts for common influences affecting all participants in a specific event, such as environmental factors, drafting regulations, group dynamics, and other shared conditions. By doing so, the standard error estimates better reflect the true underlying variation present in the data.

\subsubsection*{Two-Way Fixed Effects (TWFE) Standard Errors}

To further control for unobserved heterogeneity across both athlete and event dimensions, I employ the Two-Way Fixed Effects (TWFE) variance estimator. This approach removes systematic differences associated with individual athletes and specific events, ensuring that standard error estimates are robust to persistent unobserved factors. By eliminating these fixed effects, the estimator better isolates the true effect of interest while maintaining statistical validity.

\newpage

\section*{Results}

\subsection*{The Covid-19 Response Policy Induced Event}
During the pandemic, \textbf{individually staggered race starts} and precautionary COVID-19 measures rendered the benefits of open-water swim drafting statistically \textbf{insignificant} in 2020. In the years following the pandemic, the estimated effects of drafting appear to be somewhat less \textbf{pronounced}. This trend is found to be particularly evident in a subsample of athletes who \textbf{competed} across all time periods, potentially due to COVID-19-related health shocks, changes in training behavior, or panel data limitations such as attrition. (See Appendix)

\begin{figure}[h]  
    \centering
    \includegraphics[width=0.8\textwidth]{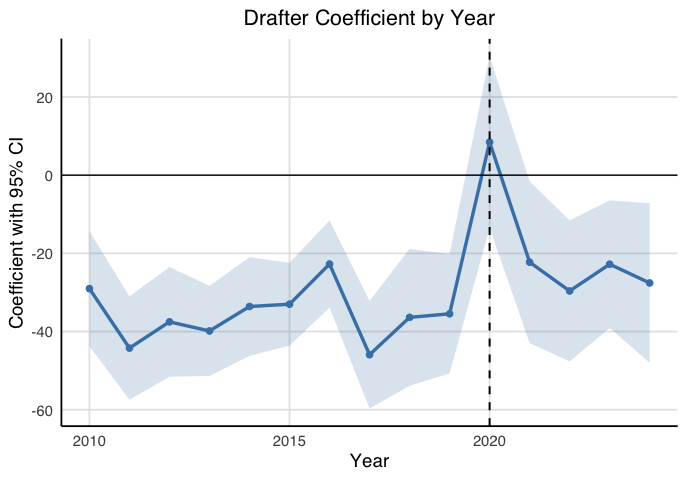}
    \caption{  
        Drafting benefits based on non-random yearly sub-samples: Coefficient Plot. The solid thin line at 0 represents  
        the baseline with no effect. The dashed vertical line at \( x = 2020 \) marks the year when mass starts were  
        disallowed at many events as a precautionary measure due to the COVID-19 pandemic. Furthermore COVID-19 related race regulations were in place. In this year the coefficient of computed drafter is found to be insignificant. 
    }

    \label{fig:coefplot}
\end{figure}

\begin{table}[ht!]
    \centering
    \begin{threeparttable}
        \caption{OLS Estimation Results - Dependent Variable: Swim Out Times}
        \label{tab:ols_Swim-Out Times1}
        \begin{tabular}{lcccc}
            \toprule
            & Estimate & Std. Error & t-value & Pr(>|t|) \\
            \midrule
            Drafter & -23.8886 & 4.1184 & -5.8004 & $6.63 \times 10^{-9}$ *** \\
            Pre-period $\times$ Drafter & -8.5779 & 4.6013 & -1.8642 & 0.0623 . \\
            Covid-year $\times$ Drafter & 19.4150 & 8.1469 & 2.3831 & 0.0172 * \\
            \midrule
            Observations & \multicolumn{4}{c}{168,391} \\
            Fixed-effects & \multicolumn{4}{c}{Athlete ID: 29,194, Event ID: 1,339, Cluster (Swim Group): 224} \\
            Standard-errors & \multicolumn{4}{c}{Heteroskedasticity-robust} \\
            RMSE & \multicolumn{4}{c}{175.1} \\
            Adj. $R^2$ & \multicolumn{4}{c}{0.9683} \\
            Within $R^2$ & \multicolumn{4}{c}{0.0035} \\
            \bottomrule
        \end{tabular}
        
        \begin{tablenotes}
            \item \textbf{Notes:}
            \item The dependent variable is Swim-Out Times performance (time in seconds). Fixed-effects regression includes athlete ID, event ID, and computed swim group cluster in event.
            \item \textit{Periods:} The variable \textit{pre-period} is a dummy indicating events before 2020. The variable \textit{covid-year} is a dummy indicating events in 2020.
            \item \textit{Swim groups:} Clusters are determined by hierarchical clustering with a threshold of 5 seconds.
            \item \textit{Dummies:} The \textit{drafter} dummy is 1 for non-leaders of swim groups. The \textit{pre-period * drafter} interaction tests the effect of drafting before 2020. The \textit{covid-year * drafter} interaction examines the drafting effect in 2020.
            \item \textit{Significance codes:} *** $p<0.001$, ** $p<0.01$, * $p<0.05$, . $p<0.1$.
        \end{tablenotes}
    \end{threeparttable}
\end{table}

\subsection*{The Effect of Higher-Order Drafting Positions} 
Drafting is \textbf{beneficial} compared to the leader (based purely on swim-out segment performance time) only when an athlete is in at least the third drafting \textbf{position} within a swim group constructed based on clustered swim-out times. Drafting positions 1 and 2 are therefore \textbf{strategic} primarily as positions to minimize \textbf{fatigue} and \textbf{strain} before the subsequent bike leg in triathlon races. \citet{Delextrat2003} Since swim groups are constructed based on swim-out times, there may be some degree of \textbf{endogeneity} in the estimates. This could arise because athletes who manage to catch up to a swim group toward the end of the swim segment might \textbf{influence} their position within the swim group, potentially leading to biased results. To mitigate some of this endogeneity, \textbf{athlete} and \textbf{event} fixed effects are employed. \\ The inversely \textbf{decaying} nature of swim drafting benefits addresses at least some possible concerns of simultaneous reverse \textbf{causality} and partially also omitted variable \textbf{bias}.
\newpage

\begin{figure}[ht!]
    \centering
    \includegraphics[width=0.49\textwidth]{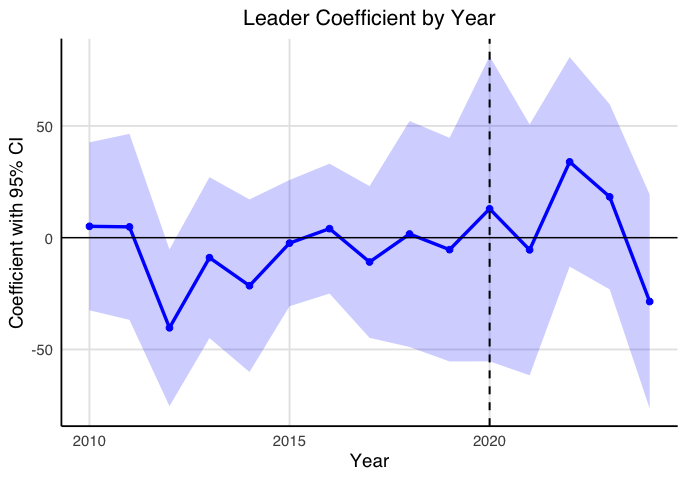}
    \includegraphics[width=0.5\textwidth]{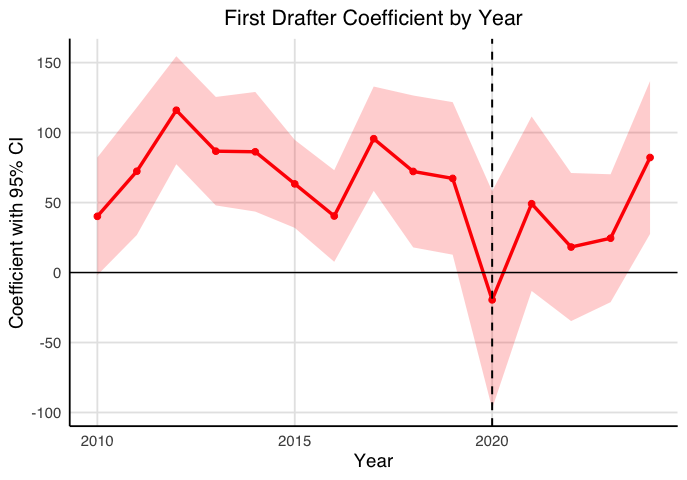}
    \includegraphics[width=0.49\textwidth]{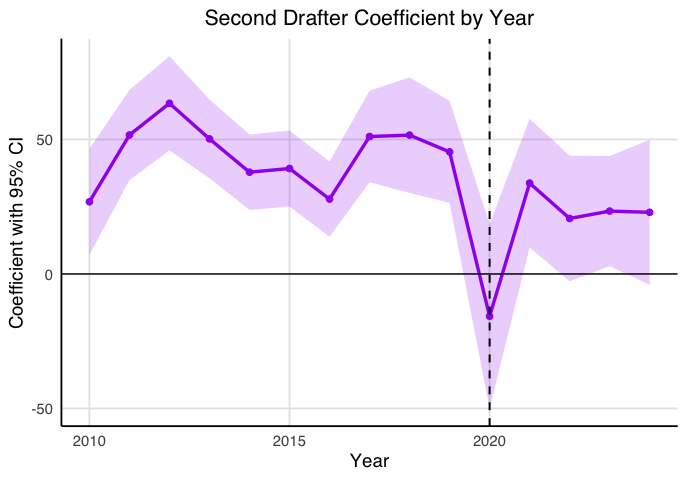}
    \includegraphics[width=0.5\textwidth]{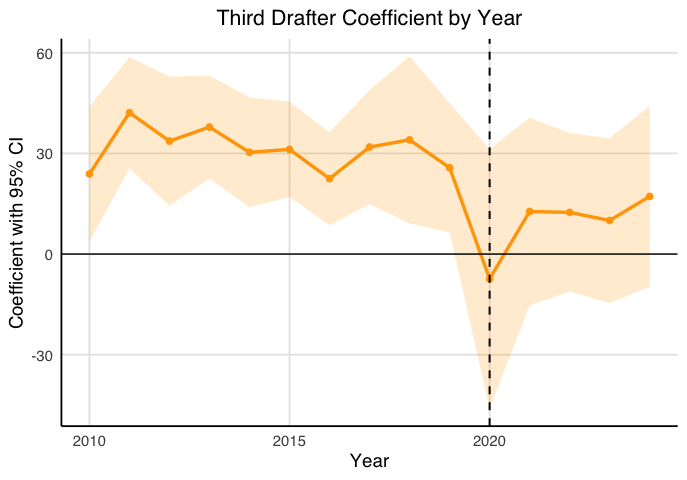}
    \includegraphics[width=0.49\textwidth]{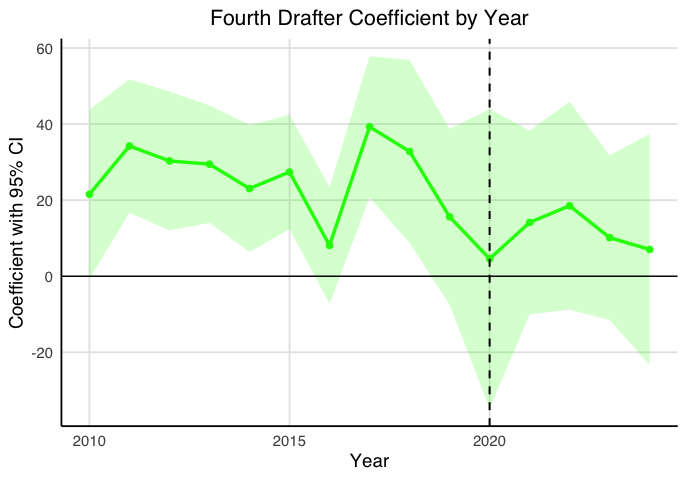}
    \includegraphics[width=0.5\textwidth]{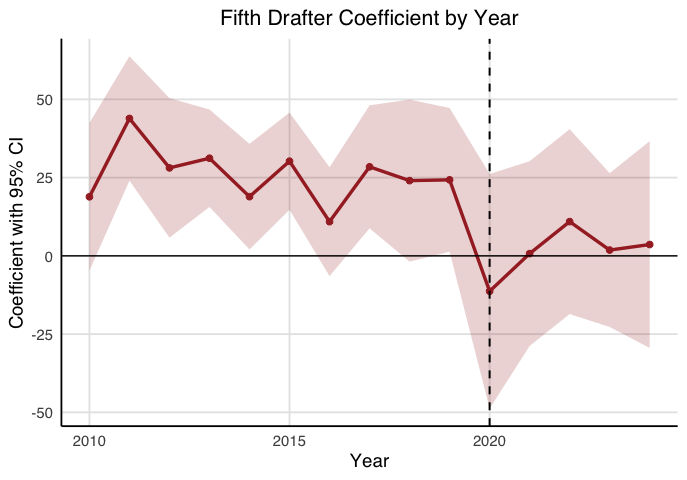}
    \caption{
        Yearly coefficient plots from fixed-effects regressions estimating the impact of drafting positions  
        and leadership on swim-out time performance. Each model is estimated separately for every year  
        from 2010 to 2024, controlling for individual athlete and event fixed effects. Notable drafting position  
        dummies are found to have no significant penalty effect in 2020, when mass starts were disallowed as a  
        precautionary measure at many events due to the COVID-19 pandemic. Higher-order drafting positions  
        are found to have less pronounced penalties in any given year, making them more favorable.}

    \label{fig:coefplots}
\end{figure}

\begin{table}[h]
    \centering
     \renewcommand{\arraystretch}{0.7}
    \begin{threeparttable}
        \caption{Combined Drafting Position Dummies FE Model Estimation Results}
        \label{tab:fe_model}
        \begin{tabular}{lcccccc}
                \toprule
                    \multicolumn{7}{c}{\textbf{\shortstack{OLS Estimation of FE Models (Dependent Variable: Swim-Out Time, \\ Swim Group Def.: Max 5 sec. apart, Cluster (Swim Group) Single Linkage)}}} \\
            
        \toprule
 
            & Model 1 & Model 2 & Model 3 & Model 4 & Model 5 & Model 6 \\
            \midrule
            \textbf{Leader}          & 0.8688  & 13.7712$^{***}$  & 33.0137$^{***}$  & 31.4543$^{***}$  & 30.8295$^{***}$  & 30.4761$^{***}$  \\
            & (3.9616)  & (3.7570)  & (1.4801)  & (1.5119)  & (1.4979)  & (1.4965)  \\
            \textbf{Leader x Cluster (SG)}  & -  & -  & -  & -  & -  & 1.2926$^{***}$  \\
            & -  & -  & -  & -  & -  & (0.1264)  \\
            \textbf{Cluster (SG)}         & -       & -  & -  & -  & -  & 0.6579$^{***}$  \\
            & -       & -  & -  & -  & -  & (0.0862)  \\
            \textbf{Race Rank}            & -       & -  & -  & -  & -  & 0.3967$^{***}$  \\
            & -       & -  & -  & -  & -  & (0.0072)  \\
            \textbf{First Drafter}   & 46.8571$^{***}$  & 17.8360$^{***}$  & -  & -  & -  & 48.5491$^{***}$  \\
            & (4.3806)  & (3.9534)  & -  & -  & -  & (4.2379)  \\
            \textbf{Second Drafter}  & 26.0330$^{***}$  & -  & 10.7271$^{***}$  & -  & -  & 31.2118$^{***}$  \\
            & (1.8769)  & -  & (1.5851)  & -  & -  & (1.7903)  \\
            \textbf{Third Drafter}   & 19.4259$^{***}$  & -  & -  & 7.1819$^{***}$  & -  & 23.5610$^{***}$  \\
            & (1.8238)  & -  & -  & (1.6871)  & -  & (1.7470)  \\
            \textbf{Fourth Drafter}  & 17.4769$^{***}$  & -  & -  & -  & 6.0963$^{**}$  & 19.6363$^{***}$  \\
            & (1.9906)  & -  & -  & -  & (1.8925)  & (1.8914)  \\
            \textbf{Fifth Drafter}   & 15.3214$^{***}$  & -  & -  & -  & -  & 15.4568$^{***}$  \\
            & (2.0454)  & -  & -  & -  & -  & (1.9386)  \\
            \midrule
            Observations   & 168,391 & 168,391 & 168,391 & 168,391 & 168,391 & 168,391 \\
            RMSE          & 175.0   & 175.1   & 175.1   & 175.1   & 175.1   & 169.2   \\
            Adj. $R^2$    & 0.9684  & 0.9683  & 0.9683  & 0.9683  & 0.9683  & 0.9705  \\
            Within $R^2$  & 0.0050  & 0.0035  & 0.0037  & 0.0035  & 0.0034  & 0.1131  \\
            \bottomrule
        \end{tabular}
        
        \begin{tablenotes}
            \item \textit{Notes:} This table presents the results of six fixed-effects OLS estimations. The dependent variable is \textit{Swim-Out Times}. Models include fixed effects for \textit{athlete\_id} (29,194 levels) and \textit{event\_id} (1,339 levels). Standard errors are robust to heteroskedasticity. 
            \item Statistical significance: *** $p<0.001$, ** $p<0.01$, * $p<0.05$.
        \end{tablenotes}
    \end{threeparttable}
\end{table}

\clearpage

\newpage

\section*{Discussion}

While I provide the first causal estimates of open-water swim drafting effects, several limitations should be acknowledged.  

First, the estimation approach relies on inferred swim group structures based on clustered Swim-Out times, which may introduce measurement error. The classification of drafters and leaders remains an approximation rather than a direct observation.  

Second, the staggered race starts during the COVID-19 pandemic serve as a natural experiment. However, unobserved factors, such as changes in athlete fitness levels, race-day conditions, or competition field strength, may have influenced drafting benefits across different periods, potentially confounding causal interpretations. This might have led to the only weak significance (at the 10\% level) of the interaction term \textit{Pre-period $\times$ Drafter} in Table \ref{tab:ols_Swim-Out Times1}.  

Third, reverse causality remains a concern despite the inclusion of athlete- and event-fixed effects. More capable athletes may self-select into different drafting positions, catch up to swim groups over the course of a race, or utilize higher passive drag depending on their fitness level on race day. These dynamics could result in endogenous group formations, potentially biasing the estimated effects.  

Finally, course-specific factors such as buoy placements, water currents, and water temperature---which influence wetsuit usage regulations---are not explicitly controlled for in the models. These factors may affect drafting effectiveness but are difficult to quantify systematically across different race events.

\section*{Conclusion}

This study provides the first large-scale causal estimate of open-water swim drafting effects in real-world triathlon race settings. Leveraging a natural experiment induced by staggered race starts during the COVID-19 pandemic, I find that drafting benefits were statistically insignificant in 2020 but re-emerged post-pandemic, albeit at slightly reduced levels. Our results indicate that drafting only becomes beneficial from the third trailing position onward, while earlier positions primarily serve to minimize fatigue or strategic purposes. By employing athlete and event fixed effects, I address potential heterogeneity concerns, and the observed inverse decay of drafting benefits helps to partially mitigate concerns related to endogeneity, reverse causality,, and omitted variable bias. These findings offer valuable insights into optimal swim strategies for endurance athletes and race organizers.

\newpage

\section*{Declaration of Interest}
The authors declare no conflicts of interest. 

\section*{Funding Information}
No funding for this research was provided.

\section*{Confirmation of Ethical Compliance}
The study adheres to ethical guidelines, and no human subjects were directly involved in the research.

\section*{CRediT Author Statement}
\textbf{Conceptualization}: Felix Reichel; \\
\textbf{Methodology}: Felix Reichel; \\
\textbf{Formal analysis}: Felix Reichel; \\
\textbf{Investigation}: Felix Reichel; \\
\textbf{Data curation}: Manuel Poeter MSc, Mag. Thomas Doblhammer; \\
\textbf{Writing – original draft}: Felix Reichel; \\
\textbf{Writing – review \& editing}: Felix Reichel; \\
\textbf{Visualization}: Felix Reichel; \\
\textbf{Supervision}: Felix Reichel.

\newpage

\bibliographystyle{chicago}

\begin{thebibliography}{}

\bibitem[Bolon et al.(2020)]{Bolon2020}
Bolon, B., Pretot, C., Clanet, C., Larrarte, F., and Carmigniani, R. (2020). Drafting of two passive swimmer scale models for open-water races. \textit{Soft Matter}, \textit{16}(31), 7270--7273. doi: \href{https://doi.org/10.1103/PhysRevFluids.00.004800}{10.1103/PhysRevFluids.00.004800}.

\bibitem[Chatard et al.(1990)]{Chatard1990}
Chatard, J.-C., Lavoie, J.-M., Bourgoin, B., and Lacour, J.-R. (1990). The contribution of passive drag as a determinant of swimming performance. \textit{International Journal of Sports Medicine}, \textit{11}(4), 367--372.

\bibitem[Chatard and Wilson(2003)]{ChatardWilson2003}
Chatard, J.-C., and Wilson, B. (2003). Drafting distance in swimming. \textit{Medicine \& Science in Sports \& Exercise}, \textit{35}(7), 1176--1180. doi: \href{https://doi.org/10.1249/01.MSS.0000074564.06106.1F}{10.1249/01.MSS.0000074564.06106.1F}.

\bibitem[Delextrat et al.(2003)]{Delextrat2003}
Delextrat, A., Tricot, V., Bernard, T., Vercruyssen, F., and Hausswirth, C. (2003). Drafting during swimming improves efficiency during subsequent cycling. \textit{Medicine \& Science in Sports \& Exercise}, \textit{35}(9), 1612--1619.

\bibitem[Leitner(2021)]{Leitner2021}
Leitner, S. (2021). Massenstart oder Einzelstart? \textit{Triathlon Szene Österreich}, January 8. Retrieved from \href{https://www.trinews.at/massenstart-oder-einzelstart/}{https://www.trinews.at/massenstart-oder-einzelstart/}.

\bibitem[Millet et al.(1999)]{Millet1999}
Millet, G., Chatard, J.-C., and Chollet, D. (1999). Effects of kicking, performance, and drag during draft swimming in elite triathletes. In \textit{Les Cahiers de l'INSEP} (No. 24, pp. 209--216). Proceedings of the 2nd INSEP International Triathlon Congress.

\bibitem[Ohmichi et al.(1983)]{Ohmichi1983}
Ohmichi, H., Takamoto, M., and Miyashita, M. (1983). Measurement of the waves caused by swimmers. In Hollander, A. P., Huijing, P. A., and de Groot, G. (Eds.), \textit{Biomechanics and Medicine in Swimming V} (pp. 103--107). Champaign, IL: Human Kinetics.

\bibitem[Olbrecht(2011)]{Olbrecht2011}
Olbrecht, J. (2011). Triathlon: Swimming for winning. \textit{Journal of Human Sport and Exercise}, \textit{6}(2), 233--246.

\bibitem[Toussaint et al.(1989)]{Toussaint1989}
Toussaint, H. M., Bruinink, R., Coster, R., and others. (1989). Effect of triathlon wet suit on drag during swimming. \textit{Medicine \& Science in Sports \& Exercise}, \textit{21}(3), 325--328.

\bibitem[Österreichischer Triathlonverband (ÖTRV)(2020)]{Triathlon2020}
Österreichischer Triathlonverband (ÖTRV). (2020). ALOHA TRI Steyregg feiert Premiere. \textit{Triathlon Austria}. Retrieved from \href{https://www.triathlon-austria.at/de/newsshow-alle-----aloha-tri-steyregg-feiert-premiere?return=240}{https://www.triathlon-austria.at/de/newsshow-alle-----aloha-tri-steyregg-feiert-premiere?return=240}.

\end{thebibliography}

\clearpage

\newpage

\subsection*{Appendices}

\section*{Using Covid-19 Response Policy to Estimate Open Water
Swim Drafting Effects in Triathlon}

\begin{landscape}
    \begin{table}[h!]
        \centering
            \renewcommand{\arraystretch}{0.8}
        \begin{threeparttable}
        \caption{OLS Regression Results Considering all Participants in Non-Leading Swim-Group Roles as Potential Drafting Beneficiaries}
        \label{tab:regression_results_vertical}
        \begin{tabular}{>{\raggedright\arraybackslash}p{5cm} >{\centering\arraybackslash}p{3.5cm} >{\centering\arraybackslash}p{3.5cm} >{\centering\arraybackslash}p{3.5cm} >{\centering\arraybackslash}p{3.5cm}}
            \toprule
            & \textbf{(1) Pre-Covid Sample} ($\leq 2020 $) & \textbf{(2) Covid Sample} (2020, 2021, 2022) & \textbf{(3) Post-Covid Sample} ($\geq 2023$) & \textbf{(4) Full Sample} (2010 - 2024) \\
            \midrule
            \multicolumn{5}{>{\centering\arraybackslash}p{19cm}}{\textbf{Panel A: OLS Estimations (Swim Group Def.: Max 5 sec apart, Cluster (Swim Group) Complete Linkage, Subsample based on Athletes: Competing in all 3 Periods)}} \\
            \midrule
            \textbf{Observations} & 29,750 & 9,796 & 6,941 & 46,487 \\
            \textbf{Fixed Effects} & Event, Athlete & Event, Athlete & Event, Athlete & Event, Athlete \\
            \textbf{Clustered SE} & Event ID & Event ID & Event ID & Event ID \\
            \textbf{Drafter} & -19.476*** & -13.061*** & -15.648** & -18.569*** \\
            \textbf{Standard Error} & (2.413) & (3.517) & (4.668) & (1.855) \\
            \textbf{RMSE} & 171.3 & 147.9 & 129.7 & 178.9 \\
            \textbf{Adjusted R$^2$} & 0.9709 & 0.9742 & 0.9793 & 0.9698 \\
            \textbf{Within R$^2$} & 0.0021 & 0.0012 & 0.0020 & 0.0019 \\
            \midrule
            \multicolumn{5}{>{\centering\arraybackslash}p{19cm}}{\textbf{Panel B: OLS Estimations (Swim Group Def.: Max 5 sec apart, Single Swimmers are allowed, Cluster (Swim Group) Complete Linkage)}} \\
            \midrule
            \textbf{Observations} & 135,505 & 18,467 & 14,419 & 168,391 \\
            \textbf{Fixed Effects} & Event, Athlete & Event, Athlete & Event, Athlete & Event, Athlete \\
            \textbf{Clustered SE} & Event ID & Event ID & Event ID & Event ID \\
            \textbf{Drafter} & -22.892*** & -18.307*** & -22.388*** & -23.047*** \\
            \textbf{Standard Error} & (1.459) & (4.179) & (4.365) & (1.330) \\
            \textbf{RMSE} & 166.6 & 140.1 & 116.1 & 179.4 \\
            \textbf{Adjusted R$^2$} & 0.9717 & 0.9707 & 0.9781 & 0.9668 \\
            \textbf{Within R$^2$} & 0.0035 & 0.0024 & 0.0045 & 0.0030 \\
            \midrule
            \multicolumn{5}{>{\centering\arraybackslash}p{19cm}}{\textbf{Panel C: OLS Estimations (Swim Group Def.: Max 15 sec apart, Cluster (Swim Group) Complete Linkage, Subsample based on Athletes: At Least 3 Swimmers per Group)}} \\
            \midrule
            \textbf{Observations} & 111,880 & 14,262 & 11,011 & 137,153 \\
            \textbf{Fixed Effects} & Event, Athlete & Event, Athlete & Event, Athlete & Event, Athlete \\
            \textbf{Clustered SE} & Event ID & Event ID & Event ID & Event ID \\
            \textbf{Drafter} & -5.348*** & -3.206 & -9.764* & -5.339*** \\
            \textbf{Standard Error} & (0.951) & (2.620) & (4.082) & (0.841) \\
            \textbf{RMSE} & 122.7 & 92.1 & 76.5 & 123.4 \\
            \textbf{Adjusted R$^2$} & 0.9809 & 0.9826 & 0.9860 & 0.9803 \\
            \textbf{Within R$^2$} & 0.00024 & 0.00013 & 0.00139 & 0.00024 \\
            \midrule
            \multicolumn{5}{l}{\textit{Significance Codes:} *** $p<0.001$, ** $p<0.01$, * $p<0.05$} \\
            \bottomrule
        \end{tabular}
        \end{threeparttable}
    \end{table}
\end{landscape}

\clearpage
\newpage

\begin{table}[htbp]
    \centering
    \footnotesize
    \renewcommand{\arraystretch}{0.7}
    \setlength{\tabcolsep}{6pt} 
    \begin{tabular}{p{4cm} p{2.25cm} p{2.25cm} p{2.25cm} p{2.25cm}} 
        \toprule
        \multicolumn{5}{c}{\textbf{\shortstack{Panel D: OLS Estimation (Dependent Variable: Swim-Out Time,
        Clustered SE at Athlete Level, \\ Swim Group Def.: Max 5 sec. apart, Cluster (Swim Group) Single Linkage)}}} \\
        \midrule
        \textbf{Observations} & \multicolumn{4}{c}{168,391} \\
        \textbf{Fixed Effects} & \multicolumn{4}{c}{Athlete (29,194), Event (1,339)} \\
        \textbf{Clustered SE} & \multicolumn{4}{c}{Athlete ID} \\
        \midrule
        \textbf{Variable} & \textbf{Estimate} & \textbf{Std. Error} & \textbf{t value} & \textbf{Pr(>|t|)} \\
        \midrule
        \textbf{Pre-period x Drafter} & -51.680*** & 1.841 & -28.065 & < 2.2e-16 \\
        \textbf{Post-period x Drafter} & 15.138*** & 3.629 & 4.171 & 3.04e-05 \\
        \midrule
        \textbf{RMSE} & \multicolumn{4}{c}{178.8} \\
        \textbf{Adjusted R$^2$} & \multicolumn{4}{c}{0.9670} \\
        \textbf{Within R$^2$} & \multicolumn{4}{c}{0.0099} \\
        \midrule
        \multicolumn{5}{c}{\textbf{\shortstack{Panel E: OLS Estimation (Dependent Variable: Swim-Out Time, Additional Controls,\\ Swim Group Def.: Max 5 sec. apart, Cluster (Swim Group) Single Linkage)}}} \\
        \midrule
        \textbf{Observations} & \multicolumn{4}{c}{168,391} \\
        \textbf{Fixed Effects} & \multicolumn{4}{c}{Athlete (29,194), Event (1,339)} \\
        \textbf{Clustered SE} & \multicolumn{4}{c}{Athlete ID} \\
        \midrule
        \textbf{Pre-period x Drafter} & -29.858*** & 1.880 & -15.886 & < 2.2e-16 \\
        \textbf{Post-period x Drafter} & 14.012*** & 3.627 & 3.864 & 0.00011 \\
        \textbf{Drafter Position} & -0.173*** & 0.018 & -9.398 & < 2.2e-16 \\
        \textbf{Cluster (Swim Group)} & 1.669*** & 0.107 & 15.564 & < 2.2e-16 \\
        \midrule
        \textbf{RMSE} & \multicolumn{4}{c}{177.4} \\
        \textbf{Adjusted R$^2$} & \multicolumn{4}{c}{0.9675} \\
        \textbf{Within R$^2$} & \multicolumn{4}{c}{0.0251} \\
        \midrule
        \multicolumn{5}{c}{\textbf{\shortstack{Panel F: OLS Estimation (Dependent Variable: Swim-Out Time, Clustered SE at Athlete \& Event Level,\\ Swim Group Def.: Max 5 sec. apart, Cluster (Swim Group) Single Linkage)}}} \\
        \midrule
        \textbf{Observations} & \multicolumn{4}{c}{168,391} \\
        \textbf{Fixed Effects} & \multicolumn{4}{c}{Athlete (29,194), Event (1,339)} \\
        \textbf{Clustered SE} & \multicolumn{4}{c}{TWFE (Athlete ID, Event ID)} \\
        \midrule
        \textbf{Pre-period x Drafter} & -29.967*** & 2.433 & -12.317 & < 2.2e-16 \\
        \textbf{Post-period x Drafter} & 12.543** & 4.357 & 2.879 & 0.0041 \\
        \textbf{Drafter Position} & -0.271** & 0.091 & -2.978 & 0.0030 \\
        \textbf{Cluster (Swim Group)} & 1.518*** & 0.124 & 12.285 & < 2.2e-16 \\
        \textbf{Race Rank} & 0.402*** & 0.030 & 13.316 & < 2.2e-16 \\
        \midrule
        \textbf{RMSE} & \multicolumn{4}{c}{169.4} \\
        \textbf{Adjusted R$^2$} & \multicolumn{4}{c}{0.9704} \\
        \textbf{Within R$^2$} & \multicolumn{4}{c}{0.1111} \\
        \midrule
        \multicolumn{5}{l}{\textit{Significance codes:} *** $p<0.001$, ** $p<0.01$, * $p<0.05$} \\
        \midrule
        \multicolumn{5}{c}{\textbf{\shortstack{Panel G: OLS Estimation (Additional Swim Group Cluster-Level Fixed Effects, Robust SE,\\ Swim Group Def.: Max 5 sec. apart, Cluster (Swim Group) Single Linkage)}}} \\
        \midrule
        \textbf{Fixed Effects} & \multicolumn{4}{c}{Cluster (Swim Group) (224), Athlete (29,194), Event (1,339)} \\
        \textbf{Clustered SE} & \multicolumn{4}{c}{Robust SE (HC1)} \\
        \midrule
        \textbf{Pre-period x Drafter} & -31.401*** & 1.652 & -19.013 & < 2.2e-16 \\
        \textbf{Post-period x Drafter} & 12.562*** & 3.594 & 3.495 & 0.00047 \\
        \textbf{Drafter Position} & -0.180*** & 0.018 & -10.277 & < 2.2e-16 \\
        \midrule
        \textbf{RMSE} & \multicolumn{4}{c}{175.0} \\
        \textbf{Adjusted R$^2$} & \multicolumn{4}{c}{0.9684} \\
        \textbf{Within R$^2$} & \multicolumn{4}{c}{0.0045} \\
        \midrule
        \multicolumn{5}{l}{\textit{Significance codes:} *** $p<0.001$, ** $p<0.01$, * $p<0.05$} \\
        \bottomrule
    \end{tabular}
    \caption{OLS Estimation Results for Swim-Out Times Under Different Model Specifications}
  \begin{minipage}{0.95\textwidth}
        \small
        \textbf{Notes:} This table presents results from OLS regressions estimating the effects of drafting on Swim-out times performance. (lower is better) 
        \textbf{Panel D} estimates the effects using an interaction terms specification, clustering at the athlete level.
        \textbf{Panel E} introduces additional controls such as the swimmer's relative position in the drafting group ($drafter\_position$) and cluster ID denoting the swim group.
        \textbf{Panel F}  further refines the estimation by clustering standard errors at both the athlete and event level. 
        The dependent variable is the \textit{Swim-Out Time} time in all models. 
        \textbf{Panel G} extends the analysis by including \textbf{cluster-level fixed effects}, which account for within-group variations in drafting performance, 
        while using HC1 standard errors robust to heteroskedasticity.

        The pre-period covers events before 2020 (2010-2019), while the post-period includes competitions during covid precautionary actions (from 2020-2023). 
        Standard errors are clustered as indicated. 
        Significance codes: *** $p\leq0.001$, ** $p\leq0.01$, * $p\leq0.05$.
    \end{minipage}
    \label{tab:ols_results}
\end{table}

\clearpage
\newpage

\newpage

\begin{table}[h]
    \centering
    \begin{threeparttable}
        \caption{OLS Estimation Results - Dependent Variable: Swim Out Times}
        \label{tab:ols_Swim-Out Times}
        \begin{tabular}{lcccc}
            \toprule
            & Estimate & Std. Error & t-value & Pr(>|t|) \\
            \midrule
            Drafter & -23.8886 & 4.1184 & -5.8004 & $6.63 \times 10^{-9}$ *** \\
            Pre-period $\times$ Drafter & -8.5779 & 4.6013 & -1.8642 & 0.0623 . \\
            Covid-year $\times$ Drafter & 19.4150 & 8.1469 & 2.3831 & 0.0172 * \\
            \midrule
            Observations & \multicolumn{4}{c}{168,391} \\
            Fixed-effects & \multicolumn{4}{c}{Athlete ID: 29,194, Event ID: 1,339, Cluster (Swim Group): 224} \\
            Standard-errors & \multicolumn{4}{c}{Heteroskedasticity-robust} \\
            RMSE & \multicolumn{4}{c}{175.1} \\
            Adj. $R^2$ & \multicolumn{4}{c}{0.9683} \\
            Within $R^2$ & \multicolumn{4}{c}{0.0035} \\
            \bottomrule
        \end{tabular}
        
        \begin{tablenotes}
            \item \textbf{Notes:}
            \item The dependent variable is Swim-Out Times performance (time in seconds). Fixed-effects regression includes athlete ID, event ID, and computed swim group cluster in event.
            \item \textit{Periods:} The variable \textit{pre-period} is a dummy indicating events before 2020. The variable \textit{covid-year} is a dummy indicating events in 2020.
            \item \textit{Swim groups:} Clusters are determined by hierarchical clustering with a threshold of 5 seconds.
            \item \textit{Dummies:} The \textit{drafter} dummy is 1 for non-leaders of swim groups. The \textit{pre-period * drafter} interaction tests the effect of drafting before 2020. The \textit{covid-year * drafter} interaction examines the drafting effect in 2020.
            \item \textit{Significance codes:} *** $p<0.001$, ** $p<0.01$, * $p<0.05$, . $p<0.1$.
        \end{tablenotes}
    \end{threeparttable}
\end{table}

\clearpage
\newpage

\begin{figure}[h]  
    \centering
    \includegraphics[width=0.9\textwidth]{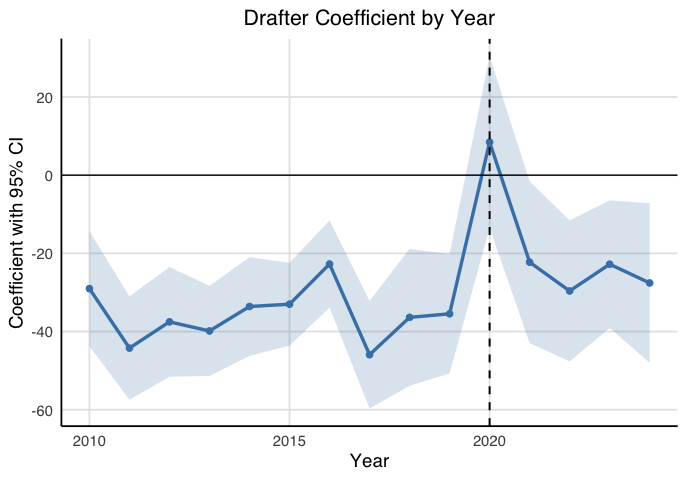}
    \caption{  
        Drafting benefits based on non-random yearly sub-samples: Coefficient Plot. The solid thin line at 0 represents  
        the baseline with no effect. The dashed vertical line at \( x = 2020 \) marks the year when mass starts were  
        disallowed at many events as a precautionary measure due to the COVID-19 pandemic. In this year the coefficient of computed drafter is found to be insignificant. 
    }

    \label{fig:coefplot}
\end{figure}

\clearpage
\newpage

\begin{figure}[h]
    \centering
    \includegraphics[width=0.9\textwidth]{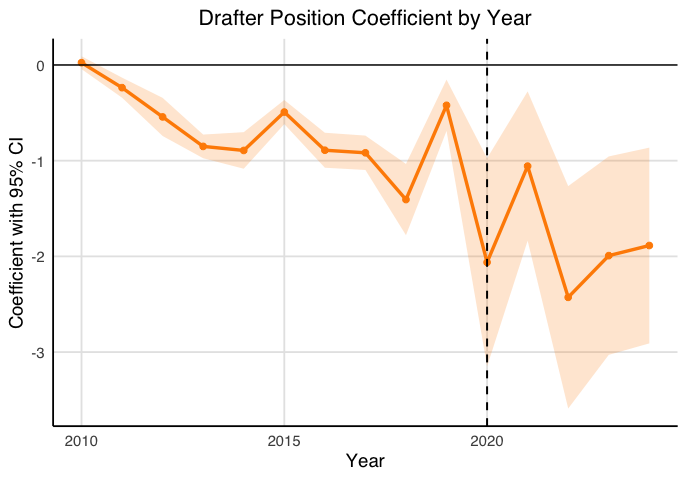}
    \caption{  
        Higher-Order Drafting Position Benefits Based on Non-Random Yearly Sub-Samples: Coefficient Plot.  
        This plot displays the estimated coefficients for drafting positions, illustrating their impact on  
        positive swim-out time performance, which corresponds to a negative effect on measured swim-out time.  
        The effect remains persistent even in the year when COVID-19 precautionary measures, such as  
        disallowing mass starts, were implemented.  
    }

    \label{fig:poscoef}
\end{figure}

\clearpage
\newpage

\newpage

\begin{table}[h]
    \centering
    \begin{threeparttable}
        \caption{Summary Statistics of numerical variables}
        \label{tab:summary_stats}
        \begin{tabular}{lcccccc}
            \toprule
            Variable & Min & 1st Quartile & Median & Mean & 3rd Quartile & Max \\
            \midrule
            Total & 165 & 4965 & 8232 & 11789 & 16809 & 61282 \\
            Swim-Out Times & 4 & 825 & 1431 & 1625 & 2097 & 29700 \\
            Race Rank & 0 & 28 & 69 & 156.9 & 151 & 2623 \\
            Male & 0.0000 & 1.0000 & 1.0000 & 0.7898 & 1.0000 & 1.0000 \\
            Year & 1922 & 1969 & 1977 & 1977 & 1985 & 2009 \\
            Covid & 0.0000 & 0.0000 & 0.0000 & 0.1097 & 0.0000 & 1.0000 \\
            Post Covid & 0.00000 & 0.00000 & 0.00000 & 0.08563 & 0.00000 & 1.00000 \\
            Week Running & 0.0 & 159.0 & 279.0 & 317.6 & 475.0 & 751.0 \\
            Age & 8.00 & 31.00 & 38.00 & 38.56 & 46.00 & 99.00 \\
            Age Squared & 64 & 961 & 1444 & 1600 & 2116 & 9801 \\
            Event Year & 2010 & 2013 & 2015 & 2016 & 2019 & 2024 \\
            Cluster (Swim Group) & 1.00 & 4.00 & 10.00 & 16.95 & 23.00 & 224.00 \\
            Leader & 0.0000 & 0.0000 & 0.0000 & 0.3134 & 1.0000 & 1.0000 \\
            Drafter & 0.0000 & 0.0000 & 1.0000 & 0.6866 & 1.0000 & 1.0000 \\
            Drafter Position & 1.00 & 1.00 & 3.00 & 17.15 & 11.00 & 937.00 \\
            First Drafter & 0.0000 & 0.0000 & 0.0000 & 0.2961 & 1.0000 & 1.0000 \\
            Second Drafter & 0.0000 & 0.0000 & 0.0000 & 0.1342 & 0.0000 & 1.0000 \\
            Third Drafter & 0.00000 & 0.00000 & 0.00000 & 0.08329 & 0.00000 & 1.00000 \\
            Fourth Drafter & 0.00000 & 0.00000 & 0.00000 & 0.05846 & 0.00000 & 1.00000 \\
            Fifth Drafter & 0.000 & 0.000 & 0.000 & 0.044 & 0.000 & 1.000 \\
            Last Drafter & 0.0000 & 0.0000 & 0.0000 & 0.2961 & 1.0000 & 1.0000 \\
            \bottomrule
        \end{tabular}
        \begin{tablenotes}
            \item \textit{Notes:} This table presents summary statistics for the key variables used in the analysis. The dataset consists of 168,391 observations. Variables include athlete characteristics, race details, and encoded swim group (cluster), drafting, drafting position binary dummy variables. Encoded categorical variables such as Event Category \(\in \{\text{Sprint}, \text{Short}, \text{Middle}, \text{Long}\} \) are not displayed.
        \end{tablenotes}
    \end{threeparttable}
\end{table}

\clearpage
\newpage

\newpage

\begin{table}[h]
    \centering
    \begin{threeparttable}
        \caption{Drafting Position Dummies FE Model Estimation Results}
        \label{tab:fe_model}
        \begin{tabular}{lcccc}
            \toprule
            & Estimate & Std. Error & t value & Pr(>|t|) \\
            \midrule
            First Drafter  & 71.9698  & 1.72555 & 41.70837 & < 2.2e-16 *** \\
            Second Drafter & 39.4901  & 1.83705 & 21.49654 & < 2.2e-16 *** \\
            Third Drafter  & 28.1694  & 1.84785 & 15.24444 & < 2.2e-16 *** \\
            Fourth Drafter & 23.5146  & 2.02665 & 11.60269 & < 2.2e-16 *** \\
            Fifth Drafter  & 19.5379  & 2.08455 & 9.37268  & < 2.2e-16 *** \\
            \bottomrule
        \end{tabular}
        
        \begin{tablenotes}
            \item \textit{Notes:} This table presents the results of an ordinary least squares (OLS) estimation with fixed effects. The dependent variable is \textit{Swim-Out Time (sec)}. The model includes fixed effects for \textit{athlete\_id} (29,194 levels) and \textit{event\_id} (1,339 levels). Robust standard errors are used to account for heteroskedasticity.
            \item The number of observations is 168,391. The root mean squared error (RMSE) is 178.5. The adjusted $R^2$ is 0.967166, and the within-group $R^2$ is 0.013724.
            \item Statistical significance codes: *** $p<0.001$, ** $p<0.01$, * $p<0.05$.
        \end{tablenotes}
    \end{threeparttable}
\end{table}

\clearpage
\newpage

\newpage

\begin{table}[h]
    \centering
    \begin{threeparttable}
        \caption{Contd. Drafting Position Dummies FE Model Estimation Results}
        \label{tab:fe_model}
        \begin{tabular}{lcccc}
            \toprule
            & Estimate & Std. Error & t value & Pr(>|t|) \\
            \midrule
            \multicolumn{5}{c}{\textbf{Panel A: OLS Estimation FE Model with Drafting Positions}} \\
            \midrule
            First Drafter  & 71.9698  & 1.72555 & 41.70837 & < 2.2e-16 *** \\
            Second Drafter & 39.4901  & 1.83705 & 21.49654 & < 2.2e-16 *** \\
            Third Drafter  & 28.1694  & 1.84785 & 15.24444 & < 2.2e-16 *** \\
            Fourth Drafter & 23.5146  & 2.02665 & 11.60269 & < 2.2e-16 *** \\
            Fifth Drafter  & 19.5379  & 2.08455 & 9.37268  & < 2.2e-16 *** \\
            \midrule
            \multicolumn{5}{c}{\textbf{Panel B: OLS Estimation FE Model with Drafting Positions and Additional Controls}} \\
            \midrule
            Leader         & -26.7843  & 4.35217  & -6.15423  & 7.5649e-10 *** \\
            Cluster (Swim Group)        & 0.65793   & 0.08622  & 7.63061   & 2.3514e-14 *** \\
            Race Rank          & 0.39668   & 0.00723  & 54.85970  & < 2.2e-16 *** \\
            First Drafter  & 48.5491   & 4.23790  & 11.45593  & < 2.2e-16 *** \\
            Second Drafter & 31.2118   & 1.79029  & 17.43395  & < 2.2e-16 *** \\
            Third Drafter  & 23.5610   & 1.74702  & 13.48643  & < 2.2e-16 *** \\
            Fourth Drafter & 19.6363   & 1.89143  & 10.38174  & < 2.2e-16 *** \\
            Fifth Drafter  & 15.4568   & 1.93864  & 7.97301   & 1.5604e-15 *** \\
            Leader x Cluster (Swim Group) & 1.29262   & 0.12643  & 10.22384  & < 2.2e-16 *** \\
            \bottomrule
        \end{tabular}
        
        \begin{tablenotes}
            \item \textit{Notes:} This table presents the results of two ordinary least squares (OLS) estimations with fixed effects. The dependent variable is \textit{Swim-Out Times}. The models include fixed effects for \textit{athlete\_id} (29,194 levels) and \textit{event\_id} (1,339 levels). Robust standard errors are used to account for heteroskedasticity.
            \item The number of observations is 168,391. Model 1 has a root mean squared error (RMSE) of 178.5, an adjusted $R^2$ of 0.967166, and a within-group $R^2$ of 0.013724. Model 2 has an RMSE of 169.2, an adjusted $R^2$ of 0.970474, and a within-group $R^2$ of 0.113118.
            \item Statistical significance codes: *** $p<0.001$, ** $p<0.01$, * $p<0.05$.
        \end{tablenotes}
    \end{threeparttable}
\end{table}

\clearpage
\newpage

\begin{table}[h]
    \centering
    \begin{threeparttable}
        \caption{Combined Drafting Position Dummies FE Model Estimation Results}
        \label{tab:fe_model}
        \begin{tabular}{lccc}
            \toprule
            & Model 1 & Model 2 & Model 3 \\
            \midrule
            \textbf{Leader}          & -       & -26.7843$^{***}$  & 0.8688  \\
            &         & (4.3522)  & (3.9616)  \\
            \textbf{Cluster (Swim Group)}         & -       & 0.6579$^{***}$  & -  \\
            &         & (0.0862)   & -  \\
            \textbf{Race Rank}            & -       & 0.3967$^{***}$  & -  \\
            &         & (0.0072)   & -  \\
            \textbf{First Drafter}   & 71.9698$^{***}$  & 48.5491$^{***}$  & 46.8571$^{***}$  \\
            & (1.7256)  & (4.2379)  & (4.3806)  \\
            \textbf{Second Drafter}  & 39.4901$^{***}$  & 31.2118$^{***}$  & 26.0330$^{***}$  \\
            & (1.8371)  & (1.7903)  & (1.8769)  \\
            \textbf{Third Drafter}   & 28.1694$^{***}$  & 23.5610$^{***}$  & 19.4259$^{***}$  \\
            & (1.8479)  & (1.7470)  & (1.8238)  \\
            \textbf{Fourth Drafter}  & 23.5146$^{***}$  & 19.6363$^{***}$  & 17.4769$^{***}$  \\
            & (2.0267)  & (1.8914)  & (1.9906)  \\
            \textbf{Fifth Drafter}   & 19.5379$^{***}$  & 15.4568$^{***}$  & 15.3214$^{***}$  \\
            & (2.0846)  & (1.9386)  & (2.0454)  \\
            \midrule
            Observations   & 168,391 & 168,391 & 168,391 \\
            RMSE          & 178.5   & 169.2   & 175.0   \\
            Adj. $R^2$    & 0.9672  & 0.9705  & 0.9684  \\
            Within $R^2$  & 0.0137  & 0.1131  & 0.0050  \\
            \bottomrule
        \end{tabular}
        
        \begin{tablenotes}
            \item \textit{Notes:} This table presents the results of three fixed-effects OLS estimations. The dependent variable is \textit{Swim-Out Times}. Models include fixed effects for \textit{athlete\_id} (29,194 levels) and \textit{event\_id} (1,339 levels). Standard errors are robust to heteroskedasticity. 
            \item Statistical significance: *** $p<0.001$, ** $p<0.01$, * $p<0.05$.
        \end{tablenotes}
    \end{threeparttable}
\end{table}

\clearpage
\newpage

\begin{table}[h]
    \centering
    \begin{threeparttable}
        \caption{Cont. Combined Drafting Position Dummies FE Model Estimation Results}
        \label{tab:fe_model}
        \begin{tabular}{lcccccc}
            \toprule
            & Model 1 & Model 2 & Model 3 & Model 4 & Model 5 & Model 6 \\
            \midrule
            \textbf{Leader}          & 0.8688  & 13.7712$^{***}$  & 33.0137$^{***}$  & 31.4543$^{***}$  & 30.8295$^{***}$  & 30.4761$^{***}$  \\
            & (3.9616)  & (3.7570)  & (1.4801)  & (1.5119)  & (1.4979)  & (1.4965)  \\
            \textbf{Leader x Cluster (SG)}  & -  & -  & -  & -  & -  & 1.2926$^{***}$  \\
            & -  & -  & -  & -  & -  & (0.1264)  \\
            \textbf{Cluster (SG)}         & -       & -  & -  & -  & -  & 0.6579$^{***}$  \\
            & -       & -  & -  & -  & -  & (0.0862)  \\
            \textbf{Race Rank}            & -       & -  & -  & -  & -  & 0.3967$^{***}$  \\
            & -       & -  & -  & -  & -  & (0.0072)  \\
            \textbf{First Drafter}   & 46.8571$^{***}$  & 17.8360$^{***}$  & -  & -  & -  & 48.5491$^{***}$  \\
            & (4.3806)  & (3.9534)  & -  & -  & -  & (4.2379)  \\
            \textbf{Second Drafter}  & 26.0330$^{***}$  & -  & 10.7271$^{***}$  & -  & -  & 31.2118$^{***}$  \\
            & (1.8769)  & -  & (1.5851)  & -  & -  & (1.7903)  \\
            \textbf{Third Drafter}   & 19.4259$^{***}$  & -  & -  & 7.1819$^{***}$  & -  & 23.5610$^{***}$  \\
            & (1.8238)  & -  & -  & (1.6871)  & -  & (1.7470)  \\
            \textbf{Fourth Drafter}  & 17.4769$^{***}$  & -  & -  & -  & 6.0963$^{**}$  & 19.6363$^{***}$  \\
            & (1.9906)  & -  & -  & -  & (1.8925)  & (1.8914)  \\
            \textbf{Fifth Drafter}   & 15.3214$^{***}$  & -  & -  & -  & -  & 15.4568$^{***}$  \\
            & (2.0454)  & -  & -  & -  & -  & (1.9386)  \\
            \midrule
            Observations   & 168,391 & 168,391 & 168,391 & 168,391 & 168,391 & 168,391 \\
            RMSE          & 175.0   & 175.1   & 175.1   & 175.1   & 175.1   & 169.2   \\
            Adj. $R^2$    & 0.9684  & 0.9683  & 0.9683  & 0.9683  & 0.9683  & 0.9705  \\
            Within $R^2$  & 0.0050  & 0.0035  & 0.0037  & 0.0035  & 0.0034  & 0.1131  \\
            \bottomrule
        \end{tabular}
        
        \begin{tablenotes}
            \item \textit{Notes:} This table presents the results of six fixed-effects OLS estimations. The dependent variable is \textit{Swim-Out Times}. Models include fixed effects for \textit{athlete\_id} (29,194 levels) and \textit{event\_id} (1,339 levels). Standard errors are robust to heteroskedasticity. 
            \item Statistical significance: *** $p<0.001$, ** $p<0.01$, * $p<0.05$.
        \end{tablenotes}
    \end{threeparttable}
\end{table}

\clearpage
\newpage

\newpage
\begin{figure}[h]
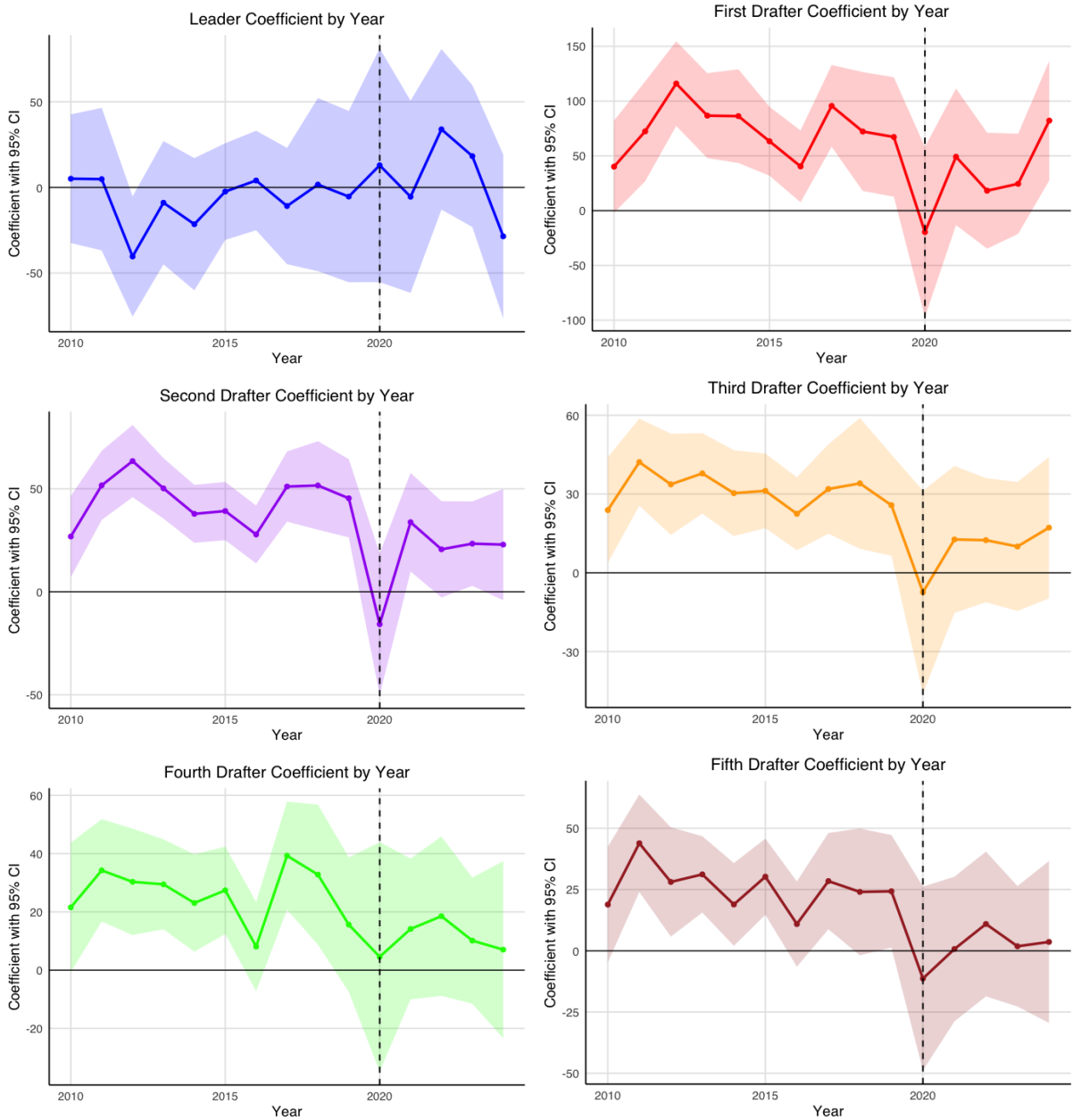

    \centering
    \includegraphics[width=0.49\textwidth]{img/leader_over_yearly.png}
    \includegraphics[width=0.5\textwidth]{img/first_drafter_over_yearly.png}
    \includegraphics[width=0.49\textwidth]{img/second_drafter_over_yearly.png}
    \includegraphics[width=0.5\textwidth]{img/third_drafter_over_yearly.png}
    \includegraphics[width=0.49\textwidth]{img/fourth_drafter_over_yearly.png}
    \includegraphics[width=0.5\textwidth]{img/fifth_drafter_over_yearly.png}
    \caption{
        Yearly coefficient plots from fixed-effects regressions estimating the impact of drafting positions  
        and leadership on swim-out time performance. Each model is estimated separately for every year  
        from 2010 to 2024, controlling for individual athlete and event fixed effects. Notable drafting position  
        dummies are found to have no significant penalty effect in 2020, when mass starts were disallowed as a  
        precautionary measure at many events due to the COVID-19 pandemic. Higher-order drafting positions  
        are found to have less pronounced penalties in any given year, making them more favorable.}

    \label{fig:coefplots}
\end{figure}

\clearpage
\newpage

    \begin{figure}[h]
        \centering
        \includegraphics[width=0.9\textwidth]{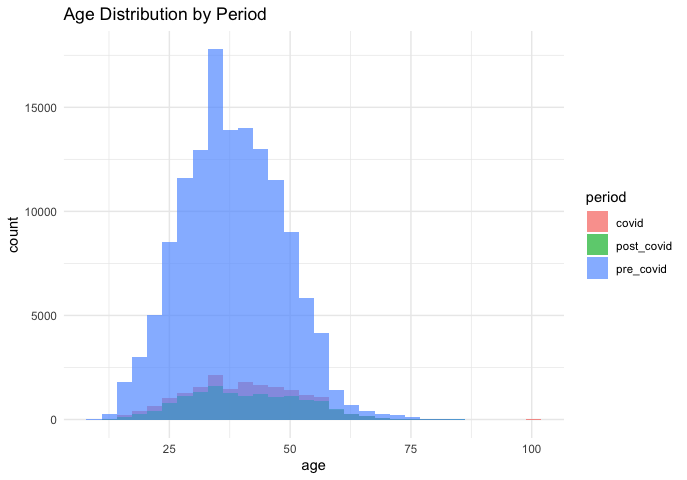}
        \caption{Age Distribution by Period Balance Check}
        \label{fig:age_dist}
    \end{figure}
    
    \begin{figure}[h]
        \centering
        \includegraphics[width=0.9\textwidth]{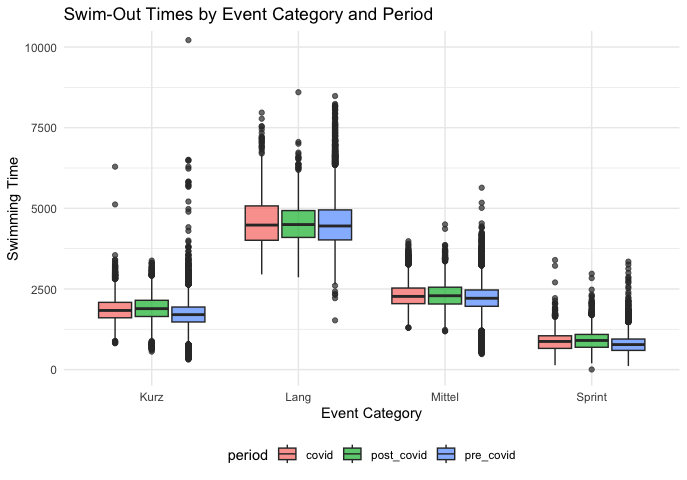}
        \caption{Swim Out Times by Event Category and Period Balance Check}
        \label{fig:swim_out_times}
    \end{figure}

\clearpage
\newpage

    \begin{table}[h]
    \centering
    \caption{Balance Check of Swim-Out Times across different Event Categories and Periods}
    \begin{tabular}{llrrrr}
        \toprule
        Event Category & Period & Mean Swim-Out Time & SD Swim-Out  Time & Count \\
        \midrule
        Short   & Covid       & 1855  & 380  & 4694  \\
        Short   & Post-Covid  & 1899  & 403  & 3726  \\
        Short   & Pre-Covid   & 1696  & 415  & 37722 \\
        \midrule
        Long   & Covid       & 4594  & 804  & 1130  \\
        Long   & Post-Covid  & 4572  & 1035 & 1014  \\
        Long   & Pre-Covid   & 4553  & 773  & 10201 \\
        \midrule
        Middle & Covid       & 2311  & 380  & 4658  \\
        Middle & Post-Covid  & 2315  & 394  & 3527  \\
        Middle & Pre-Covid   & 2216  & 436  & 26711 \\
        \midrule
        Sprint & Covid       & 854   & 295  & 7985  \\
        Sprint & Post-Covid  & 906   & 296  & 6152  \\
        Sprint & Pre-Covid   & 774   & 258  & 60871 \\
        \bottomrule
    \end{tabular}
\end{table}

\begin{table}[h]
    \centering
    \caption{Balance Check across periods}
    \begin{tabular}{lrrrrrr}
        \toprule
        Period     & Mean Total & Mean Swimming Time & Mean Rank & Mean Age & Mean Male & Count \\
        \midrule
        Covid       & 12091  & 1705  & 121  & 40.2  & 0.768  & 18467  \\
        Post-Covid  & 12573  & 1765  & 127  & 40.8  & 0.757  & 14419  \\
        Pre-Covid   & 11665  & 1599  & 165  & 38.1  & 0.796  & 135505 \\
        \bottomrule
    \end{tabular}
\end{table}

\end{document}